\documentstyle[12pt,epsf]{article}
 \def\tstrut{\vrule height2.5ex depth0pt width0pt} % used in tables
 \def\jtstrut{\vrule height5ex depth0pt width0pt} % used in tables
\begin{document}
 \begin{titlepage}
 \thispagestyle{empty}
 \begin{flushright}
 UG-DFM-1/2000 \\
  nucl-th/0001060
 \end{flushright}  
   \vspace*{5mm}

 \vspace*{2cm}

 \begin{center}
 {\Large \bf Deeply bound levels in kaonic atoms.}

 \vspace{1.5cm}
 {\large{\bf A. Baca, C. Garc\'\i a-Recio and J.
 Nieves}}\\[2em]
 Departamento de F\'{\i}sica Moderna, Universidad de Granada, 
 E-18071 Granada, Spain.

 \end{center}

 \vspace{2cm} \begin{abstract} Using a microscopic antikaon-nucleus
 optical potential recently developed by Ramos and Oset~\cite{RO00}
 from a chiral model, we calculate strong interaction energy shifts
 and widths for $K^-$ atoms. This purely theoretical potential gives
 an acceptable description of the measured data ($\chi^2/{\rm
 num. data}= 3.8$), though it turns out to be less attractive than
 what can be inferred from the existing kaon atomic data. We also use
 a modified potential, obtained by adding to the latter theoretical
 one a $s$-wave term which is fitted to known experimental kaonic data
 ($\chi^2/{\rm degree~of~freedom}= 1.6$), to predict deeply bound
 $K^-$ atomic levels, not yet detected. This improved potential
 predicts, in general, states even narrower than those recently
 reported by Friedman and Gal~\cite{FG99}.  This reinforces the idea
 that these deeply atomic states can be detected and resolved by using
 suitable nuclear reactions. Besides, we also study $K^-$ and $\bar
 K^0$ nuclear bound states and compute binding energies and widths,
 for both species of antikaons, in $^{12}$C,~$^{40}$Ca and
 $^{208}$Pb. Despite of restricting our study only to potentials
 obtained from best fits to the known kaonic atom data, the dynamics
 of these nuclear bound states depends dramatically on the particular
 optical potential used.

\vspace{1cm}

 \end{abstract}

 \end{titlepage}

 \newpage

 \setcounter{page}{1}

 \section{Introduction}

	 Strong interaction shifts and widths of bound levels of
 hadronic atoms provide valuable information on the hadron-nucleus
 interaction at threshold~\cite{BF97}. We call deeply
 bound hadronic states those that cannot be detected by means of
 spectroscopic tools ($i.e.$ analyzing the emitted $X-$rays when the
 hadron decays from one atomic level to another one in the
 electromagnetic  cascade).
 This happens for low-lying levels where the overlap between the
 hadron  wave-function and the nucleus is appreciable, and as a 
 consequence the width due to strong interaction of the hadron with
 the nucleus is much larger than the electromagnetic width of the
 level. In these circumstances, the emission of $X-$rays is highly
 suppressed as compared to the hadron absorption by the
 nucleus. Indeed, the total decay width of the last observable level,
 using $X-$ray spectroscopy techniques,
 is two or three orders of magnitude larger than the upper one.
 Thus, the width of these deeply bound states are expected
 to be very large. If the half-width of a given level is equal or
 larger than the separation to the next level, then that state cannot
 be resolved. So only {\bf narrow} deeply bound  states can be
 resolved and detected. It is out of any doubt that the precise
 experimental determination of binding energies and widths of these
 states would provide a valuable insight into the complex dynamics of
 the antikaon-nucleon and antikaon-nucleus systems. Similar studies
 have been carried out for the pion case, where narrow deeply bound
 pionic atom states were predicted~\cite{fs85}-\cite{nog93}, and have
 been recently detected using nuclear reactions ~\cite{ya96}.

    By using $X-$ray spectroscopy techniques, energy shifts and widths
    of kaonic atoms levels have been measured through the whole
    periodic table\footnote{The dynamics of all these levels are
    greatly dominated by pure electromagnetic interactions and they do
    not correspond to what we call {\it deeply bound} states.}. A
    compilation of data can be found in Refs.~\cite{BF97}
    and~\cite{FG94}. Some $K^--$nucleus optical potentials have been
    successfully fitted to experimental data~\cite{BF97},
    ~\cite{FG94}-\cite{FG98} and recently used by Friedman and Gal to
    predict the existence of {\bf narrow} deeply bound levels in
    kaonic atoms~\cite{FG99}. These authors find that the $K^-$
    deeply bound atomic levels are generally narrow, with widths
    ranging for $50$ to $1500$ KeV over the entire periodic table, and
    are not very sensitive to the different density dependence of the
    $K^--$nucleus optical potentials that were used. Besides, they
    also note that, due to the strong attraction of the
    antikaon-nucleus optical potential, there must exist {\it nuclear}
    kaonic bound levels, deeper than the atomic states, which, in
    contrast, are expected to be very sensitive to the density
    dependence of the different optical potentials. 
    
    From the microscopical point of view, recently Ramos and
    Oset~\cite{RO00} have developed an optical potential for the $K^-$
    meson in nuclear matter in a self-consistent microscopic
    manner. This approach uses a $s$-wave $\bar K N$ interaction
    obtained by solving a coupled-channel Lippmann-Schwinger
    equation\footnote{This approach is inspired in the pioneering
    works of Refs.~\cite{ka95}-\cite{ka97}. Similar extensions have
    been developed in the meson-meson sector~\cite{O97}-\cite{ej99}.},
    in the $S=-1$ strangeness sector, with a kernel determined by the
    lowest-order meson-baryon chiral Lagrangian~\cite{RO97}. Though, a
    three-momentum cut-off, which breaks manifest Lorentz covariance,
    is introduced to regularize the chiral loops, the approach
    followed by the authors of~\cite{RO97} restores exact unitarity
    and it is able to accommodate the resonance $\Lambda(1405)$.
    Self-consistency turns out to be a crucial ingredient to derive
    the $K^- -$nucleus potential in Ref.~\cite{RO00} and leads to an
    optical potential considerably more shallow than those found in 
    Refs. ~\cite{BF97}, ~\cite{FG94}-\cite{FG98}. This was firstly pointed
    out by Lutz in Ref.~\cite{Lu98}, where however, $\eta$Y channels
    were not included when solving the coupled-channel
    Lippmann-Schwinger equation for the $\bar K N$ interaction, in the
    free space.  Another recent work~\cite{Sch99}, where the medium
    modifications of antikaons in dense matter are studied in a
    coupled channel calculation, for scenarios more closely related to
    the environment encountered in heavy-ion collisions, also confirms
    the importance of self-consistency to find a similarly shallow
    potential. The depth of the real potential in the interior of
    nuclei is a topic of current interest in connection with possible
    kaon condensation in astrophysical scenarios~\cite{kn86}.

 In this work we have three aims. First, to see how this new
 microscopical optical potential~\cite{RO00}, hereafter called
 $V_{\rm opt}^{(1)}$, describes the known kaonic atom levels and, if
 possible, quantify its quality.  Second, taking $V_{\rm opt}^{(1)}$
 as starting point, without modification and also adding to it a
 phenomenological part fitted to the kaonic atom data, to give
 predictions of deeply bound kaonic atoms levels and compare them with
 previous predictions. Third, to calculate the binding energies and
 widths of the nuclear kaonic states for both $K^-$ and $\bar K^0$.

 In Ref.~\cite{zaki00}, some results obtained with the potential
 $V_{\rm opt}^{(1)}$ have been very recently reported. Here, we try
 to quantify the deficiencies of this interaction and give 
 more reliable predictions for states not yet observed, thanks to the
 use of a potential which describes better the measured data. As a
 matter of example, the nuclear $K^-$ widths predicted in
 Ref.~\cite{zaki00} are about a factor two bigger than those obtained
 in this work. Thus, all nuclear states given in Ref.~\cite{zaki00} do
 have overlap with the continuum, their meaning thus being unclear.
 Furthermore, we also present here calculations for the $\bar
 K^0-$nuclear states, not mentioned in Ref.~\cite{zaki00} at all.

 The set of experimental data we have used through this work contains
 63 energy shifts and widths of kaonic atom levels. Those are all data
 compiled in Refs.~\cite{BF97} and~\cite{FG94} except for the $3d$
 energy shift in Helium and the $6h$ energy shift in
 Ytterbium~\footnote{ We exclude the first because it is too light to
 use models based on nuclear matter and the second because there are
 certain problems on its correct interpretation~\cite{BF97}.}. We
 define the strong shift for each level by means of
\begin{equation}
\varepsilon = B_c-B
\end{equation}
where  $B$ and $B_c$ are the total and the
purely electromagnetic binding energies, respectively. Both of them are
negative, and thus a negative (positive) energy shift means
that the net effect of the strong potential is repulsive (attractive). To
 compute de $K^--$nucleus bound states, we solve the Klein-Gordon 
 equation (KGE) with an electromagnetic, $V_c$, and a strong optical, $V_{\rm
 opt}$, potentials, it reads:
 $$
 \left ( -\nabla^2 + \mu^2 +  2 \mu V_{\rm opt}(r) \right ) \Psi = \left
 ( E-V_c(r)\right)^2 \Psi
 $$
 where $\mu$ is the $K^--$nucleus reduced mass, the real part of $E$  
 is the total meson energy, including its mass, and the imaginary part
 of $E$, changed of sign, is the
 half-width $\Gamma/2$ of the state. Different models
 for $V_{\rm opt}$ will be discussed in detail in the next section,
 whereas for $V_c(r)$ we use the Coulomb interaction taking exactly the
 finite-size distribution of the nucleus and adding to it the
 vacuum-polarization corrections~\cite{iz69}.

 Both the minimization numerical algorithm and the  one used to solve  
 the KGE in coordinate space, have been extensively tested in the
 similar problem of pionic atoms~\cite{nog93}.

 \section{ Experimental kaonic atom data and the  
 optical potentials} 

 The $K^--$nucleus optical potential for kaonic atoms, $V_{\rm
 opt}$, is related to the $K^--$self-energy, $\Pi_{K^-},$ inside
 of a nuclear medium. This relation reads:
 \begin{eqnarray}
 \label{eq:V1} 
 2 \mu V_{\rm opt}(r) & = &  \Pi_{K^-}(q_0=m_{K},\vec q=0, \rho_p(r),
 \rho_n(r)), 
 \end{eqnarray} 
 where the $K^--$self-energy is evaluated at threshold, and
 $\rho_{p(n)}(r)$ is the proton (neutron) density. Neglecting isovector
 effects, as we will do in the rest of the paper, the optical potential
 only depends on $\rho(r) = \rho_p(r) + \rho_n(r)$.  Charge
 densities are
 taken from~\cite{Ja74}. For each nucleus, we take the neutron matter density 
 approximately equal to the charge one, though we consider small changes,
 inspired by Hartree-Fock calculations with the DME (density-matrix
 expansion)~\cite{Ne75}. In Table~\ref{tab:dens} we compile the
 densities used through this work.\footnote{We use modified 
 harmonic oscillator (two-parameter
 Fermi)-type densities for light (medium and heavy) nuclei.}   
 However charge (neutron matter) densities do
 not correspond to proton (neutron) ones because of the finite size of
 the proton (neutron). We take that into account following the lines of
 Ref.~\cite{nog93}.

 \begin{table}
 \begin{center}
 \begin{tabular}{llll}\hline\tstrut 
 %\hline\tstrut
 Nucleus & $R_{p}$ [fm]&  $R_{n} $[fm]   & $a$ [fm]$^*$  
  \\\hline
 \tstrut 
 $^{7}$Li$^{\rm MHO}$
 &1.770 & 1.770 &0.327 \\
 $^{9}$Be$^{\rm MHO}$
 &1.791 & 1.791 & 0.611\\
 $^{10}$B$^{\rm MHO}$
 &1.710 & 1.710 & 0.837 \\
 $^{11}$B$^{\rm MHO}$
 &1.690 & 1.690 & 0.811 \\
 $^{12}$C$^{\rm MHO}$
 &1.672 & 1.672 & 1.150 \\
 $^{16}$O$^{\rm MHO}$
 &1.833 & 1.833 &1.544\\
 $^{24}$Mg
 &2.980 & 2.930 &0.551\\
 $^{27}$Al
 &2.840& 2.840 & 0.569\\
 $^{28}$Si
 &2.930 & 2.860 & 0.569\\
 $^{31}$P
 &3.078& 3.078&0.569 \\
 $^{32}$S
 &3.165 & 3.079 & 0.569\\ 
 $^{35}$Cl
 &3.395 & 3.395 & 0.569\\
 $^{40}$Ca
 &3.51 & 3.43 & 0.563\\
 $^{59}$Co 
 &4.080 & 4.180 & 0.569\\
 $^{58}$Ni
 &4.090 & 4.140 & 0.569\\ 
 $^{63}$Cu
 &4.200 & 4.296 & 0.569\\
 $^{108}$Ag
 &5.300 & 5.500 & 0.532\\ 
 $^{112}$Cd
 &5.380 & 5.580 & 0.532\\ 
 $^{115}$In
 &5.357 & 5.557 & 0.563 \\
 $^{118}$Sn
 &5.300 & 5.550 & 0.583\\
 $^{165}$Ho
 &6.180 & 6.430 & 0.570 \\
 $^{173}$Yb
 &6.280 & 6.530 & 0.610 \\
 $^{\rm nat}$Ta
 & 6.380 & 6.630 & 0.640 \\
 $^{208}$Pb
 &6.624 & 6.890 & 0.549 \\
 $^{238}$U
 &6.805 & 7.055 & 0.605\\
 \hline
 \end{tabular}
 \end{center}
 \caption{ \small Charge ($R_p, a$) and neutron matter ($R_n, a$) density
 parameters. For light nuclei ($A \le 16$) we use a modified harmonic
 oscillator (MHO) whereas for heavier nuclei ($A > 16$) a two-parameter Fermi
 distribution was used. (*) The parameter $a$ is dimensionless for the
 MHO density form.}
 \label{tab:dens}
 \end{table}
 As we mentioned in the introduction, the authors of Ref.~\cite{RO00}
 have developed an optical potential for the $K^-$ meson in nuclear
 matter in a self-consistent microscopic manner. It is based on their
 previous work \cite{RO97} on the $s$-wave meson-baryon dynamics in
 the $S=-1$ strangeness sector. There, and starting from the lowest-order
 meson-baryon chiral Lagrangian, a non-perturbative resummation of the
 bubbles in the $s$-channel\footnote{Here $s$ refers to the Mandelstam
 variable $E_{CM}^2$} is performed. Such a
 resummation leads to an exact restoration of unitarity. The model
 reproduces successfully the $\Lambda (1405)$ resonance and the $K^- p
 \to K^- p, \bar{K}^0 n, \pi^0 \Lambda, \pi^0 \Sigma, \pi^+ \Sigma^-,
 \pi^- \Sigma^+$ cross sections at low energies.  The results in
 nuclear matter are translated to finite nuclei by means of the local
 density approximation, which turns out to be exact for
 zero-range interactions~\cite{gso89}, which is the case of the
 $s$-wave part of the $K^--$nucleus optical potential for kaonic atoms.

 \begin{table}
 \begin{center}
 \begin{tabular}{lc|llll|l}\hline\tstrut
  \ &  &\multicolumn{5}{c} {Lower level energy shifts $\varepsilon$~[KeV]}\\
 \hline\tstrut
 Nucleus & ~$n l$~ & ~~~~(1)   & ~~(1m)   & ~~~~(2) & ~(2DD)  & ~~~~~~~Exp. 
  \\\hline\tstrut 
 $^{7}$Li 
  & 2p
		 &$-$0.010   &$-$0.003  &$-$0.007 &$-$0.015  &\phantom{$-$}0.002 $\pm$ 0.026\\
 $^{9}$Be 
  & 2p
		 &$-$0.068   &$-$0.039  &$-$0.060  &$-$0.077  & $-$0.079 $\pm$ 0.021\\
 $^{10}$B
  & 2p
		 &$-$0.217   &$-$0.159  &$-$0.198  &$-$0.195  & $-$0.208 $\pm$ 0.035\\
 $^{11}$B
  & 2p
		 &$-$0.235   &$-$0.199  &$-$0.209  &$-$0.182  & $-$0.167 $\pm$ 0.035\\
 $^{12}$C
  & 2p           &$-$0.623   &$-$0.630  &$-$0.556  &$-$0.481  &  $-$0.590 $\pm$ 0.080\\

 $^{16}$O
  & 3d           &$-$0.001   &\phantom{$-$}0.0001  &$-$0.0004  &$-$0.001  &  $-$0.025 $\pm$ 0.018\\

 $^{24}$Mg
  & 3d           &$-$0.057   &$-$0.028  &$-$0.034  &$-$0.036  &  $-$0.027 $\pm$ 0.015\\

 $^{27}$Al
  & 3d           &$-$0.109   &$-$0.069  &$-$0.064  &$-$0.076  &  $-$0.080 $\pm$ 0.013\\

 $^{28}$Si
  & 3d           &$-$0.192   &$-$0.135  &$-$0.118  &$-$0.145  &  $-$0.139 $\pm$ 0.014\\

 $^{31}$P
  & 3d           &$-$0.379   &$-$0.323  &$-$0.253  &$-$0.322  &  $-$0.330 $\pm$ 0.080\\

 $^{32}$S
  & 3d           &$-$0.606   &$-$0.548  &$-$0.418  &$-$0.537  &  $-$0.494 $\pm$ 0.038\\

 $^{35}$Cl
  &  3d          &$-$1.14   &$-$1.11  &$-$0.848  &$-$1.05 &  $-$1.00 $\pm$ 0.17\\

 $^{59}$Co 
  &  4f          &$-$0.185   &$-$0.14  &$-$0.105  &$-$0.152  &  $-$0.099 $\pm$ 0.106\\

 $^{58}$Ni
  &  4f          &$-$0.239   &$-$0.185  &$-$0.137  &$-$0.197  &  $-$0.223 $\pm$ 0.042\\

 $^{63}$Cu
  &  4f          &$-$0.384   &$-$0.335  &$-$0.239  &$-$0.329  &  $-$0.370 $\pm$ 0.047\\

 $^{108}$Ag
  & 5g           &$-$0.39   &$-$0.354  &$-$0.263  &$-$0.342  &  $-$0.18 $\pm$ 0.12\\

 $^{112}$Cd
  & 5g           &$-$0.53   &$-$0.49  &$-$0.37  &$-$0.46  &  $-$0.40 $\pm$ 0.10\\

 $^{115}$In
  & 5g           &$-$0.70   &$-$0.64  &$-$0.46  &$-$0.58  &  $-$0.53 $\pm$ 0.15\\

 $^{118}$Sn
  & 5g            &$-$0.89   &$-$0.82  &$-$0.57  &$-$0.71  &  $-$0.41 $\pm$ 0.18\\

 $^{165}$Ho
  & 6h           &$-$0.34   &$-$0.29  &$-$0.20  &$-$0.26  &  $-$0.30 $\pm$ 0.13\\
 $^{\rm nat}$Ta
  & 6h           &$-$1.30   &$-$1.10  &$-$0.77  &$-$0.96  &  $-$0.27 $\pm$ 0.50\\

 $^{208}$Pb
  & 7i           &$-$0.050   &$-$0.035  &$-$0.021  &$-$0.032  &  $-$0.020 $\pm$ 0.012\\

 $^{238}$U
  & 7i           &$-$0.33   &$-$0.27  &$-$0.15  &$-$0.22  &  $-$0.26 $\pm$ 0.4\\

 \hline\tstrut
 $\chi^2/dof$ & 
		 & ~~3.76  & ~~1.57 & ~~2.15  & ~~1.83  &  \\\hline
 %\jtstrut
 \end{tabular}\\
 \end{center}
 \caption{ \small
 Energy shifts of different kaonic atom levels. The
 last column corresponds to experimental data, collected in
 Refs.~\protect\cite{BF97} and~\protect\cite{FG94}. The other columns
 correspond to results using the (1), (1m), (2) and (2DD) optical
 potentials described in the text. In the
 last row we give $\chi^2$ per
 degree of freedom including the 63 experimental data of
 Tables~\protect\ref{tab:exp1}-\protect\ref{tab:exp3}.}
 \label{tab:exp1}
 \end{table}

\begin{table}
\begin{center}
\begin{tabular}{lc|llll|l}\hline\tstrut
 \  &\multicolumn{6}{c} {Lower level energy widths $\Gamma$~[KeV]}\\
\hline\tstrut
Nucleus & ~$n l$~ & ~(1)   & (1m)   & ~~(2) & (2DD)~  & ~~~~~~Exp. 
 \\\hline\tstrut 
$^{7}$Li 
 & 2p
                & 0.039   & 0.041 &0.048 &0.044  & 0.055 $\pm$ 0.029\\
$^{9}$Be 
 & 2p
                & 0.199  &0.242  &0.235  &0.189  & 0.172 $\pm$ 0.058\\
$^{10}$B
 & 2p
                & 0.551  &0.742  &0.605  &0.505  & 0.810 $\pm$ 0.100\\
$^{11}$B
 & 2p
                & 0.555  &0.787  &0.589  &0.576  & 0.700 $\pm$ 0.080\\
$^{12}$C
 & 2p           & 1.290  &1.876  &1.341  &1.396  & 1.730 $\pm$ 0.150\\

$^{16}$O
 & 3d           & 0.007  &0.006  &0.008  &0.007  & 0.017 $\pm$ 0.014\\

$^{24}$Mg
 & 3d           & 0.208  &0.237  &0.242  &0.237  & 0.214 $\pm$ 0.015\\

$^{27}$Al
 & 3d           & 0.368  &0.438  &0.427  &0.459  & 0.443 $\pm$ 0.022\\

$^{28}$Si
 & 3d           & 0.607  &0.735  &0.706  &0.762  & 0.801 $\pm$ 0.032\\

$^{31}$P
 & 3d           & 1.051  &1.287  &1.233  &1.321  & 1.440 $\pm$ 0.120\\

$^{32}$S
 & 3d           & 1.587  &1.944  &1.872  &1.987  & 2.187 $\pm$ 0.103\\

$^{35}$Cl
 & 3d           & 2.61  &3.11  &3.13  &3.29  & 2.91 $\pm$ 0.24\\

$^{59}$Co 
 & 4f           & 0.60  &0.69  &0.71  &0.73  & 0.64 $\pm$ 0.25\\

$^{58}$Ni
 & 4f           & 0.77  &0.89  &0.90  &0.94  & 1.03 $\pm$ 0.12\\

$^{63}$Cu
 & 4f           & 1.12  &1.29  &1.33  &1.34  & 1.37 $\pm$ 0.17\\

$^{108}$Ag
 & 5g           & 1.12  &1.27  &1.31  &1.32  & 1.54 $\pm$ 0.58\\

$^{112}$Cd
 & 5g           & 1.44  &1.61  &1.68  &1.69  & 2.01 $\pm$ 0.44\\

$^{115}$In
 & 5g           & 1.89  &2.05  &2.27  &2.25  & 2.38 $\pm$ 0.57\\

$^{118}$Sn
 & 5g           & 2.38  &2.52  &2.90  &2.86  & 3.18 $\pm$ 0.64\\

$^{165}$Ho
 & 6h           & 1.05  &1.11  &1.26  &1.26  & 2.14 $\pm$ 0.31\\

$^{173}$Yb
 & 6h           & 2.01  &2.06  &2.49  &2.47  & 2.39 $\pm$ 0.30\\

$^{\rm nat}$Ta
 & 6h           & 3.56  &3.56  &4.50  &4.42  & 3.76 $\pm$ 1.15\\

$^{208}$Pb
 & 7i           & 0.19  &0.21  &0.23  &0.23  & 0.37 $\pm$ 0.15\\

$^{238}$U
 & 7i           & 1.09  &1.10  &1.32  &1.33  & 1.50 $\pm$ 0.75\\
\hline\tstrut
$\chi^2/dof$ & 
 
                & 3.76  & 1.57 & 2.15  & 1.83  &  \\\hline
\jtstrut
\end{tabular}\\
\end{center}
\caption{ \small Widths of different kaonic atom levels.  Meaning of
columns and $\chi^2 /dof$ as in Table~\protect\ref{tab:exp1}. }
\label{tab:exp2} 
\end{table}

\begin{table}
\begin{center}
\begin{tabular}{lc|llll|l}\hline\tstrut
 \  &\multicolumn{6}{c} {Upper level energy widths $\Gamma$~[eV]}\\
\hline\tstrut
Nucleus & ~$n l$~ & ~(1)   & (1m)~   & ~(2) & (2DD)  & ~~~~~~Exp. 
 \\\hline\tstrut 
$^{9}$Be 
 & 3d
                & 0.02  &0.02  &0.03  &0.04  & 0.04 $\pm$ 0.02\\
$^{12}$C
 & 3d           & 0.46  &0.40  &0.55  &0.62  & 0.99 $\pm$ 0.20\\

$^{24}$Mg
 & 4f           & 0.14  &0.11  &0.17  &0.18  & 0.08 $\pm$ 0.03\\

$^{27}$Al
 & 4f           & 0.33  &0.27  &0.39  &0.38  & 0.30 $\pm$ 0.04\\

$^{28}$Si
 & 4f           & 0.67  &0.55  &0.79  &0.76  &  0.53 $\pm$ 0.06\\

$^{31}$P
 & 4f           & 1.54  &1.31  &1.81  &1.75  &  1.89 $\pm$ 0.30\\

$^{32}$S
 & 4f           & 2.82  &2.44  &3.32  &3.21  &  3.03 $\pm$ 0.29\\

$^{35}$Cl
 & 4f           & 6.3  &5.7  &7.4  &7.2  &  5.8 $\pm$ 1.7\\

$^{58}$Ni
 & 5g           & 2.4  &2.1  &2.8  &2.8  &  5.9 $\pm$ 2.3\\

$^{63}$Cu
 & 5g           & 4.09  &3.69  &4.77  &4.88  &  5.25 $\pm$ 1.06\\

$^{108}$Ag
 & 6h           & 7.4  &7.2  &8.5  &9.1  &  7.3 $\pm$ 4.7\\

$^{112}$Cd
 & 6h           & 10.4  &10.4  &12.0  &12.8  &  6.2 $\pm$ 2.8\\

$^{115}$In
 & 6h           & 15.3  &15.0  &17.8  &18.6  &  11.4 $\pm$ 3.7\\

$^{118}$Sn
 & 6h           & 21.2  &20.5  &24.8  &25.5  &  15.1 $\pm$ 4.4\\

$^{208}$Pb
 & 8j           &  1.7 &1.6  &2.0  &2.1  &  4.1 $\pm$ 2.0\\

$^{238}$U
 & 8j           &  16 &15  &19  &19  &  45 $\pm$ 24\\

\hline\tstrut
$\chi^2/dof$ & 
                & 3.76  & 1.57 & 2.15  & 1.83  &  \\\hline
\jtstrut
\end{tabular}\\
\end{center}
\caption{ \small
Widths of different kaonic atom levels. 
Meaning of columns and $\chi^2 /dof$ as in
Table~\protect\ref{tab:exp1}. Notice that units here are eV.}
\label{tab:exp3}
\end{table}

 Firstly, we consider the antikaon-selfenergy as given in
 Ref.~\cite{RO00}, and use it to define, through Eq.~(\ref{eq:V1}),
 what we call $V_{\rm opt}^{\rm (1)}$. This potential does not have
 any free parameter, all the needed input is fixed either from studies
 of meson-baryon scattering in the vacuum or from previous studies of
 pionic atoms~\cite{nog93}, and thus it is a purely theoretical
 potential. It consists of $s$-wave and $p$-wave contributions. The
 $s$-wave part is quite complete, however the $p$-wave one is far from
 complete and only contains the contributions of $\Lambda$-hole and
 $\Sigma$-hole excitations at first order. For high-lying measured
 kaonic states, calculations of energies and widths with and without
 this $p$-wave piece turn out to be essentially undistinguishable,
 thus in what follows we will ignore the $p$-wave contribution to
 $V_{\rm opt}^{\rm (1)}$. We use $V_{\rm opt}^{\rm (1)}$ to compute
 the 63 shifts and widths of the considered set of data. The obtained
 $\chi^2$ per number of data is 3.8, indicating that the agreement is
 fairly good, taking into account that the potential has no free
 parameters.  To better quantify its goodness, we also construct a
 modified optical potential, which we call $ V_{\rm opt}^{\rm (1m)}$,
 by adding to $ V_{\rm opt}^{\rm (1)}$ a phenomenological part linear
 in density, $\delta V^{\rm fit}$, characterized by a complex constant
 $\delta b_0$ as follows 
\begin{eqnarray} \label{eq:V1m} V_{\rm
 opt}^{\rm (1m)} & = & V_{\rm opt}^{\rm (1)} + \delta V^{\rm fit}\\ 
2 \mu ~\delta V^{\rm fit}(r) & = & -4\pi~(1+\frac{\mu}{m})~ \delta b_0
 ~\rho(r) , \label{eq:dV} 
\end{eqnarray}
where $m$ is the nucleon mass.  
We determine the unknown parameter $\delta b_0$ from a best
 fit to the previous set of shifts and widths of kaonic atom data,
 this yields 
\begin{eqnarray} 
\delta b_0 & = & [~(0.078\pm
 0.009)~+~i~ (-0.25\pm 0.01)~]~ {\rm fm}.  \label{eq:db0}
 \end{eqnarray} 
and the corresponding $\chi^2$ per degree of freedom
 of the best fit is $\chi^2/dof$ = 1.6. The errors on $\delta b_0$ are
 just statistical and have been obtained by increasing the value of
 $\chi^2$ by one unit.

 As a reference, we also compare these results with the ones obtained
 from a phenomenological $t\rho-$type potential suggested in
 Refs.~\cite{FG94} and~\cite{FG99}, let us
 call it $V_{\rm opt}^{\rm (2)}(r)$:
 \begin{eqnarray} 
 2 \mu ~V_{\rm opt}^{\rm (2)}(r) & = & -4\pi~(1+\frac{\mu}{m})~ b_0~ \rho(r).
 \label{eq:V2}
 \end{eqnarray} 
 By fitting the complex parameter $b_0$ in $V_{\rm opt}^{\rm (2)}$ 
 to the same set of data, we obtain
 \begin{eqnarray}
  b_0 & = & [~(0.52\pm 0.03)~+~i ~(0.80\pm 0.03)~]~ {\rm fm } \nonumber \\
 \chi^2/dof & = & 2.15
 \label{eq:b0}
 \end{eqnarray}
 This result is slightly different from the one given in
  Ref.~\cite{BF97}:  $b_0  =  [~ (0.62\pm 0.05)~ +~i ~(0.92\pm
 0.05)~]$~fm, because the  
 nuclear-matter densities used in both works are not exactly the same.

 Now, by comparing $\delta V^{\rm fit}$ with $V_{\rm opt}^{(2)}$ we
 have a rough estimate of how 
 far is the theoretical potential $V_{\rm opt}^{(1)}$
 from the empirical one $V_{\rm opt}^{(2)}$. If we compare the result for
 $\delta b_0$ in Eq.~(\ref{eq:db0}) to the one for $b_0$ in
 Eq.~(\ref{eq:b0}), we get
\begin{eqnarray}
 \frac{{\rm Re}(\delta b_0)}{{\rm Re}( b_0)} &= &~0.15 \nonumber\\
 \frac{{\rm Im}(\delta b_0)}{{\rm Im} (b_0)} &= &-0.32,
 \label{eq:ratio} 
\end{eqnarray} 
 which tells us that $\delta b_0$ is substantially smaller than $b_0$
 and hence the theoretical potential, $V_{\rm opt}^{(1)}$, gives the
 bulk of the fitted (phenomenological) potential $V_{\rm
 opt}^{(2)}$. Of course, this is only true in the range of low
 densities which are relevant for the measured atomic levels (see the
 discussion at the end of this section). Besides,
 the microscopical potential ($V_{\rm opt}^{(1)}$) needs, in order to
 provide a better agreement to data, to have a larger attractive real
 part, and a smaller absorptive imaginary part. By looking at
 Eq.~(\ref{eq:ratio}) one might quantify these deficiencies by about
 15\% and 30\% for the real and imaginary parts respectively. Taking
 into account that the imaginary part of an optical potential provides
 an effective repulsion~\cite{FG99} and that from the above discussion
 the real part of $V_{\rm opt}^{(1)}$ is not as attractive as it
 should be, it is clear that $V_{\rm opt}^{(1)}$ is less attractive
 than what can be inferred from the existing kaonic atom data. However
 it is interesting to note, that despite of this deficiency of
 attraction, the purely theoretical potential of Ref.~\cite{RO00}
 provides an acceptable description of the data, which can be better
 quantified attending to the value of 3.8 obtained for $\chi^2$ per
 number of data, quoted above, and/or looking at the results in
 Tables~\ref{tab:exp1}--\ref{tab:exp3}. In these tables, we give the
 results obtained with this potential for shifts and widths and
 compare them to the experiment and to results computed with different
 phenomenological potentials obtained from best fits to the data. As can be
 seen in the tables, the theoretical potential, $ V_{\rm
 opt}^{\rm(1)} $, quite often predicts too repulsive shifts, 
 and for the {\it lower} states it generally predicts too small widths.

 In the next sections we will analyze the predictions (energies and
 widths) of different density dependent optical potentials, all of
 them describing the known kaonic atom data, for deeply bound atomic
 as well as nuclear kaonic states . Thus it is worthwhile to consider
 also the following density dependent potential, $ V_{\rm opt}^{\rm
 (2DD)} $, 
 \begin{eqnarray} 2 \mu~ V_{\rm opt}^{\rm (2DD)}(r) & = &
 -4\pi~(1+\frac{\mu}{m}) ~\rho(r) \left ( b_0^{\rm exp} + B_0 \left (
 \frac{\rho(r)}{\rho_0}\right ) ^\alpha\right ), \label{eq:V2DD}
 \end{eqnarray} 
 used in previous studies~\cite{FG94}
 and~\cite{FG99}. In these references, $\rho_0$ is set to 0.16
 fm$^{-3}$ and the complex parameter $b_0^{\rm exp}$ is fixed
 according to the low density limit. Using empirical scattering
 lengths, it is set to 
 \begin{eqnarray} b_0^{\rm exp} & = & (-0.15+i
 0.62)~ {\rm fm}. \label{eq:b0exp} 
\end{eqnarray} 
The above  potential has three free real parameters to be adjusted: real and
 imaginary parts of the complex parameter $B_0$ and the real parameter
 $\alpha$.  A best fit gives 
\begin{eqnarray} 
 B_0 & = & [~(1.62\pm
 0.04)~+~i~ (-0.028\pm 0.009)~]~ {\rm fm}\nonumber\\ 
\alpha & = &
 0.273 \pm 0.018\nonumber\\ 
\chi^2/dof & = & 1.83 \label{eq:b0B0}
\end{eqnarray} 
 For the sake of completeness and for a better
 comparison between potentials, we present in
 Tables~\ref{tab:exp1}--~\ref{tab:exp3} the whole set of experimental
 shifts and widths data used in the fits, together with the results
 and $\chi^2/dof$  from each of the considered 
 potentials: $V_{\rm opt}^{\rm (1)}$,
 $V_{\rm opt}^{\rm (1m)}$, $V_{\rm opt}^{\rm (2)}$ and $V_{\rm
 opt}^{\rm (2DD)}$. The potential $V_{\rm opt}^{\rm (1m)}$, mostly
 based on the theoretical input of Ref.~\cite{RO00}, gives the best
 description (smallest $\chi^2/dof$) of the data. 
  
\begin{figure}
\vspace{-1.cm}
\begin{center}                                                                
\leavevmode
\makebox[0cm]{
\epsfysize = 300pt
\epsfbox{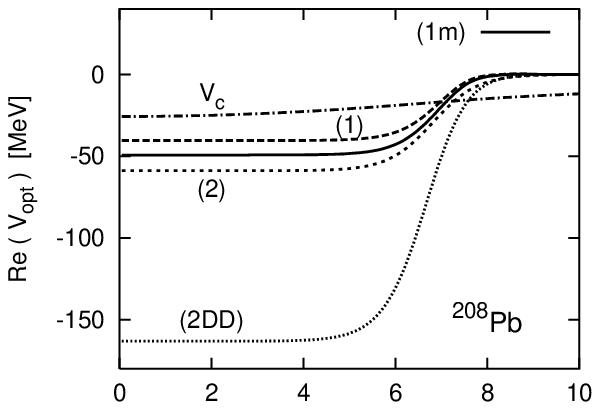}}\\ 
\vspace*{.1cm} 
\makebox[0cm]{
\epsfysize = 300pt
\epsfbox{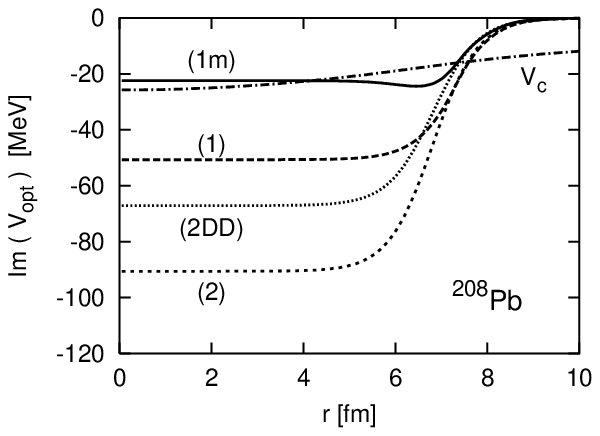}}
\end{center}
%\vspace{-15cm}
\caption {\small Real and imaginary parts of the four optical
potentials used ($V_{\rm opt}^{\rm (1)}$, $V_{\rm opt}^{\rm (1m)}$,
$V_{\rm opt}^{\rm (2)}$ and $V_{\rm opt}^{\rm (2DD)}$) in
$^{208}$Pb. For comparison the electromagnetic potential $V_c$ (real)
is also shown in both plots.}
\label{fig:Vopt1}
\end{figure}

 On the other hand, in Fig.~\ref{fig:Vopt1} we show for $^{208}$Pb,
 both the real and the imaginary parts of the four potentials studied
 in this work as a function of $r$.  In both plots, and as reference,
 the electromagnetic potential, $V_c$, felt by the $K^-$ meson is also
 depicted. Comparisons between the four potentials are of prime
 importance, because the depth of the real potential in the interior
 of nuclei determines possible kaon condensation scenarios. Another
 important feature is the shape of the different potentials in
 relation to the nuclear density. This information can also be
 extracted from Fig.~\ref{fig:Vopt1} just by noting that the
 $V^{(2)}_{\rm opt}$
 potential is proportional to $\rho$ and it is also shown in the
 plots.

\begin{figure}
\vspace{-1.cm}
\begin{center}                                                                
\leavevmode
\makebox[0cm]{
\epsfysize = 300pt
\epsfbox{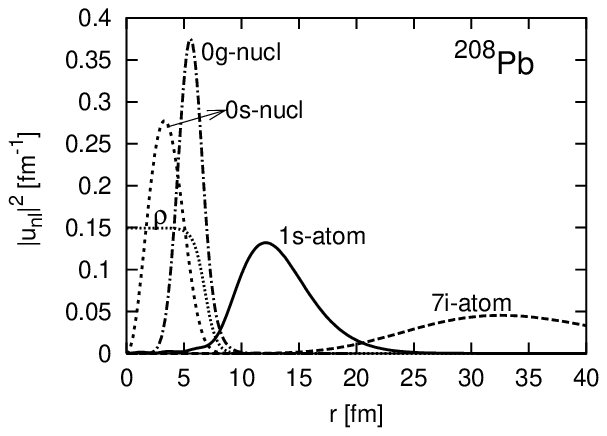}}\\ 
\vspace*{.1cm} 
\makebox[0cm]{
\epsfysize = 290pt
\epsfbox{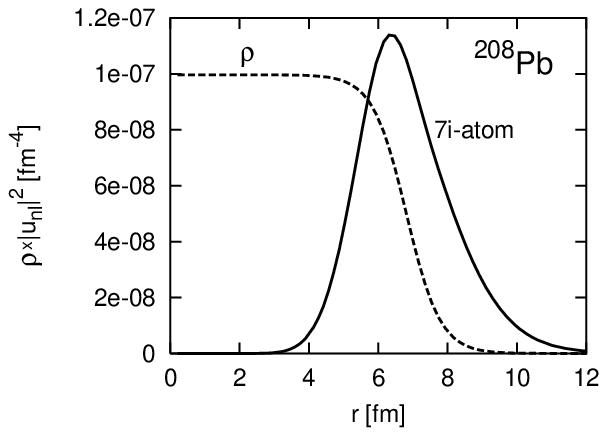}}
\end{center}
\vspace{-0.5cm}
\caption {\small Top panel: Modulus squared of the $K^--^{208}$Pb
reduced radial wave functions for the 7i, the 1s atomic
(Sect.~\protect\ref{sec:atom}) and the 0s and 0g nuclear
(Sect.~\protect\ref{sec:nucl}) levels as functions of $r$. Radial wave
functions satisfy $\int_0^\infty dr~ |u_{nl}(r)|^2 = 1$. We also show
the nuclear density $\rho$ (fm$^{-3}$) in lead. Bottom panel: The
product $\rho(r) \times |u_{nl}(r)|^2$ as a function of $r$. Here
again the nuclear density, now  in arbitrary units, is also plotted. All
wave--functions have been obtained using $V_{\rm opt}^{\rm (1m)}$.}
\label{fig:pb7i}
\end{figure}

\begin{figure}
\vspace{-1.cm}
\begin{center}                                                                
\leavevmode
\makebox[0cm]{
\epsfysize = 300pt
\epsfbox{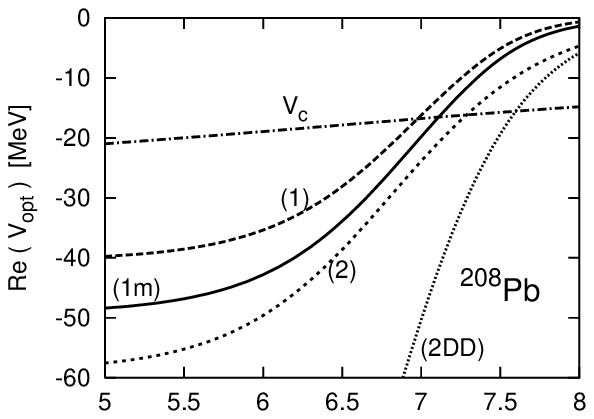}}\\ 
\vspace*{.1cm} 
\makebox[0cm]{
\epsfysize = 300pt
\epsfbox{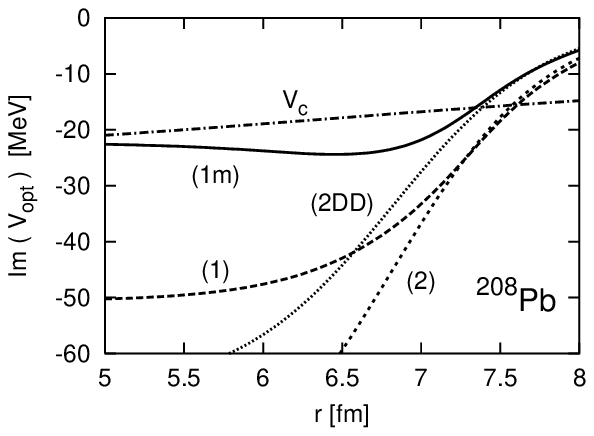}}
\end{center}
%\vspace{-15cm}
\caption {\small  Same as in Fig.\protect\ref{fig:Vopt1}.}
\label{fig:Vopt2}
\end{figure}

Despite of the fact that, for instance, $V_{\rm opt}^{\rm (1m)}$ and
$V_{\rm opt}^{\rm (2DD)}$ have a totally different depths for
distances smaller than 5 fm, both potentials give a good description
of the measured atomic levels. This is a clear indication that these
measured levels should be sensitive to low values of the nuclear
density, as can be appreciated in
Fig.~\ref{fig:pb7i}. There\footnote{In this figure, we also present
wave function for deeper bound states, not yet detected, which will be
discussed in the next sections.}, we show the $7$i modulus squared of
the $K^--^{208}$Pb reduced radial wave function, $u_{{\rm 7i}}(r)$,
and compare it with the $^{208}$Pb nuclear density ($\rho$). Though
the maximum of $|u_{{\rm 7i}}(r)|^2$ is above 30 fm, the overlap of
the kaon with the nuclear density reaches its maximum around 6 or 7 fm
(see bottom plot). Thus, the atomic data are not sensitive to the
values of the optical potentials at the center of the nuclei, but
rather to their behavior at the surface. Thus, we consider of interest
to amplify the region between 5 and 8 fm in Fig.~\ref{fig:Vopt1}. This
can be found in Fig.~\ref{fig:Vopt2}, where we see that at the surface
all optical potentials are much more similar than at the center of the
nucleus. We would like to point out that the two potentials which
better described the existing data, $V_{\rm opt}^{\rm (1m)}$ and
$V_{\rm opt}^{\rm (2DD)}$, have practically the same imaginary part
above 7 fm.

 To finish this section we would like to stress that the strong
 interaction effects in kaonic atoms are highly non-perturbative, as
 evidenced by the fact that although the level shifts are repulsive
 ({\it i.e.} negative) the real part of the optical potentials is 
 attractive. This is a direct consequence of the absorptive part of the  
 potentials being comparable in magnitude to the real part~\cite{FG94}.
\section{Deeply bound atomic $K^-$-nucleus levels}
\label{sec:atom}
We have used the four previous $K^-$-nucleus optical potentials  
($V_{\rm opt}^{\rm (1)}$, $V_{\rm opt}^{\rm (1m)}$, $V_{\rm opt}^{\rm
(2)}$ and $V_{\rm opt}^{\rm (2DD)}$)
to predict binding energies and widths of deeply
bound atomic states, not yet observed, in $^{12}{\rm C}$, $^{40}{\rm
Ca}$ and $^{208}{\rm Pb}$.  Results for binding energies, strong
shifts and widths are collected in Table~\ref{tab:atom}. Binding
energies, $B$, are practically independent of the used optical
potential. Indeed, for a given nucleus and level, $B$ varies at most
at the level of one per cent, hence we only present results obtained
with the modified Oset-Ramos potential, $V_{\rm opt}^{\rm
(1m)}$. Deeply bound atomic level widths are much more sensitive to
the details of the potential, and approximately follow a regular
pattern: $V_{\rm opt}^{\rm (2)}-$widths are the widest, those
calculated with the $V_{\rm opt}^{\rm (2DD)}$ potential are narrower
than the first ones by a few per cent, as it was previously stated in
Ref.~\cite{FG00}. For the deepest states, the $V_{\rm opt}^{\rm (1m)}$
interaction leads to significantly smaller widths (about 20 to 40 \%)
than the $V_{\rm opt}^{\rm (2)}$ potential. On the other hand, $V_{\rm
opt}^{\rm (1m)}$ and $V_{\rm opt}^{\rm (2DD)}$ predictions are more
similar, but one still finds appreciable differences. The
narrowest widths are obtained with the $V_{\rm opt}^{\rm (1)}$
potential. 

{\small 
\begin{table}
\begin{center}
\begin{tabular}{l|c|ll|llll}\hline\tstrut
 \ Nucleus &Atomic Level & $B$~[MeV] & 
$\varepsilon$~[MeV] &\multicolumn{4} {c} 
{$\Gamma$~[MeV] } \\
\hline\tstrut
& $n l$ & ~~~~(1m) & ~~(1m)   & ~(1)   & (1m) & ~(2)  & (2DD) 
 \\\hline\tstrut 
$^{12}$C 
 & 1s    &$-$0.274  &$-$0.158   &0.049  &0.036  & 0.064 & 0.055 \\
 & 2s    &$-$0.087  &$-$0.024   &0.009  &0.007  & 0.012 & 0.010 \\
 & 3s   &$-$0.042  &$-$0.008   &0.003  &0.002  & 0.004 &  0.003\\
 & 2p   &$-$0.113  &$-$0.001   &0.001  &0.002  & 0.001 &  0.001
 \\\hline\tstrut 
$^{40}$Ca 
 & 1s   &$-$1.48  &$-$1.961   &0.374  &0.456  & 0.469 & 0.414 \\
 & 2s   &$-$0.62  &$-$0.436   &0.105  &0.128  & 0.131 & 0.117 \\
 & 3s   &$-$0.34  &$-$0.161   &0.043  &0.053  & 0.054 & 0.048 \\
 & 2p   &$-$1.09  &$-$0.202   &0.133  &0.127  &0.171  & 0.164 \\
 & 3p   &$-$0.50  &$-$0.069   &0.046  &0.044  &0.059 & 0.057 \\
 & 4p   &$-$0.29  &$-$0.031   &0.021  &0.020  &0.027  & 0.026 \\
 & 5p   &$-$0.19  &$-$0.016   &0.011  &0.011  &0.014  & 0.014 \\
 & 3d   &$-$0.58  &$-$0.004   &0.007  &0.007  &0.008  & 0.008 \\
 & 4d   &$-$0.32  &$-$0.002   &0.004  &0.004  &0.005  & 0.004 \\
 & 5d   &$-$0.21  &$-$0.001   &0.002  &0.002  &0.003  & 0.003 
 \\\hline\tstrut 
$^{208}$Pb 
 & 1s   &$-$7.20  &$-$10.5   &1.39  &1.42  &1.75  &  1.60\\
 & 2s   &$-$4.19  &$-$5.20   &0.656  &0.657  &0.818  &0.749  \\
 & 3s   &$-$2.79  &$-$2.70   &0.367  &0.366  &0.457  &0.418  \\
 & 4s   &$-$2.00  &$-$1.55   &0.227  &0.226  &0.283  &0.258  \\
 & 2p   &$-$6.75  &$-$6.02   &1.20  &1.38  &1.49  &1.38  \\
 & 3p   &$-$3.96  &$-$2.88   &0.574  &0.661  &0.714  & 0.663 \\
 & 4p   &$-$2.65  &$-$1.57   &0.326  &0.373  &0.404  & 0.376 \\
 & 5p   &$-$1.91  &$-$0.939   &0.203  &0.232  &0.252  &0.235  \\
 & 3d   &$-$5.87  &$-$2.70   &0.833  &0.848  &1.04  & 0.959 \\
 & 4d   &$-$3.56  &$-$1.38   &0.424  &0.432  &0.529  & 0.486 \\
 & 5d   &$-$2.44  &$-$0.792   &0.249  &0.254  &0.310  & 0.285 \\
 & 4f   &$-$4.68  &$-$0.796   &0.418  &0.468  &0.522  & 0.492 \\
 & 5f   &$-$2.99  &$-$0.502   &0.245  &0.271  &0.305  & 0.288 \\
 & 6f   &$-$2.11  &$-$0.321   &0.154  &0.168  &0.191  & 0.181 \\
 & 5g   &$-$3.46  &$-$0.098   &0.109  &0.118  &0.136  & 0.129 \\
 & 6g   &$-$2.38  &$-$0.087   &0.086  &0.095  &0.108  & 0.102 \\
 & 7g   &$-$1.75  &$-$0.066   &0.062  &0.069  &0.077  & 0.073 \\
 & 6h   &$-$2.47  &$-$0.004   &0.009  &0.009  &0.011  & 0.011 \\
 & 7h   &$-$1.81  &$-$0.005   &0.010  &0.010  &0.013  & 0.013 \\
 & 8h   &$-$1.39  &$-$0.005   &0.009  &0.009  &0.011  & 0.011 
\\\hline

%\jtstrut
\end{tabular}\\
\end{center}
\caption{ \small Total binding energies $B$, strong shifts
$\varepsilon$,  and widths $\Gamma$ of
different deeply bound kaonic atom levels. Results for shifts and
bindings are only presented for the potential $V_{\rm opt}^{\rm
(1m)}$, whereas for the widths we give the results obtained with the four
optical potentials considered in this work: $V_{\rm opt}^{\rm (1)}$,
$V_{\rm opt}^{\rm (1m)}$, $V_{\rm opt}^{\rm (2)}$ and $V_{\rm
opt}^{\rm (2DD)}$.  }
\label{tab:atom} 
\end{table}
}

\begin{figure}
\vspace{-3.cm}
\begin{center}                                                                
\leavevmode
\epsfysize = 1050pt 
\makebox[0cm]{\epsfbox{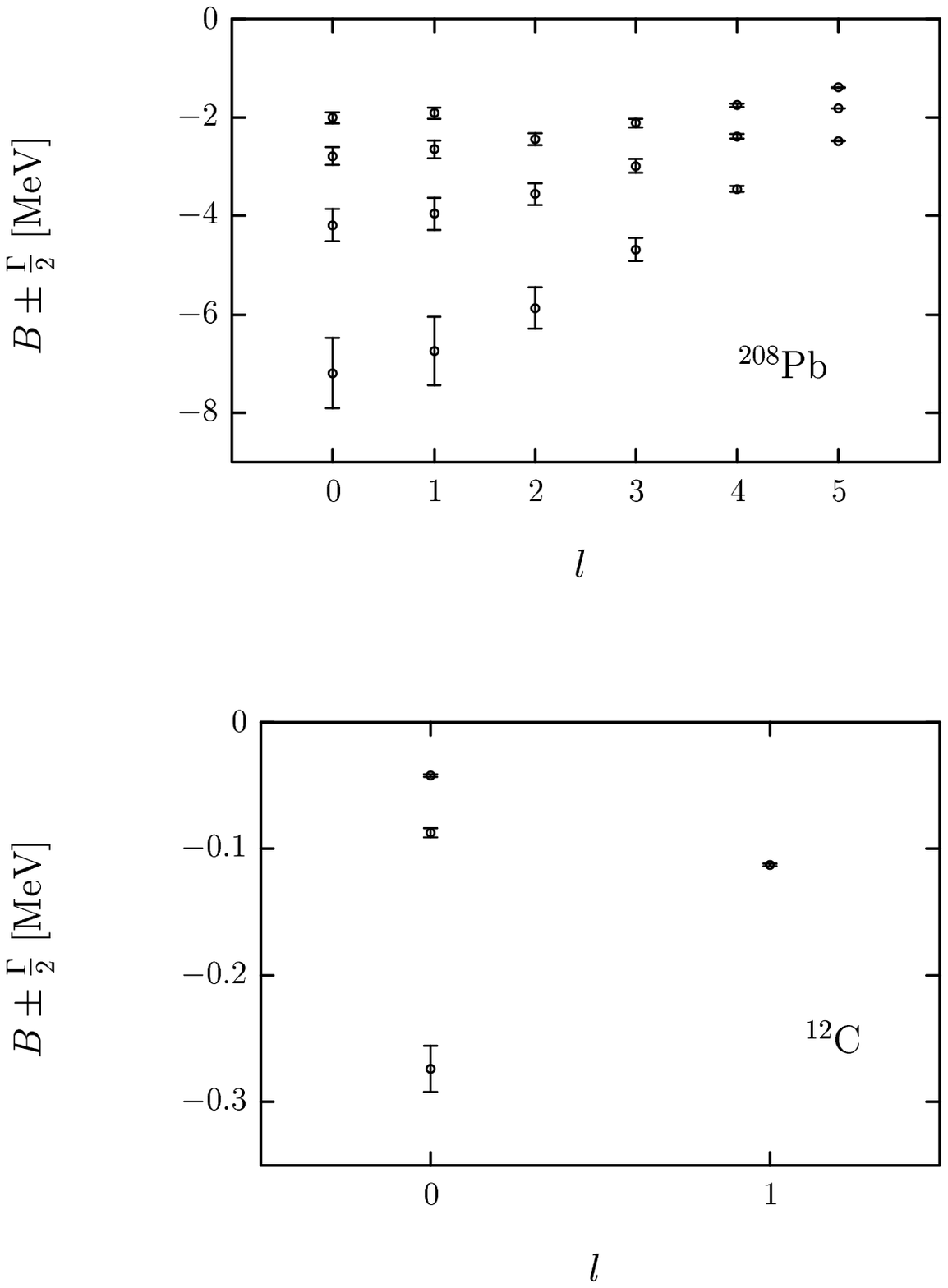}}
\end{center}
\vspace{-15cm}
\caption {\small Binding energies $B $ of deeply bound atomic levels in
$^{12}$C and $^{208}$Pb. The
error bar stands for the full width $\Gamma$ of each level. They have
been computed using the $V^{\rm(1m)}_{\rm opt}$  potential.}
\label{fig:atom}
\end{figure}

In Refs.~\cite{BF97} and~ \cite{FG99}, and supported by the results
obtained from the empirical $V_{\rm opt}^{\rm (2)}$ and $V_{\rm
opt}^{\rm (2DD)}$ potentials, a scenario where the deep atomic
states are narrow enough to be separated in most cases, except for
some overlap in heavy nuclei\footnote{In those cases, $l$-selective
nuclear reactions might resolve the whole spectrum.}, was firstly
presented. The above discussion and the results presented in 
Table~\ref{tab:atom} for the more theoretical founded potentials
$V_{\rm opt}^{\rm (1)}$ and $V_{\rm opt}^{\rm (1m)}$ reinforce such a
scenario.  To illustrate this point, we present in
Fig.~\ref{fig:atom} binding energies and widths, using the
$V_{\rm opt}^{\rm (1m)}$ potential, for different
atomic states in carbon and lead to show how separable the different
levels are.

\begin{figure}
\vspace{-1.cm}
\begin{center}                                                                
\leavevmode
\makebox[0cm]{
\epsfysize = 300pt
\epsfbox{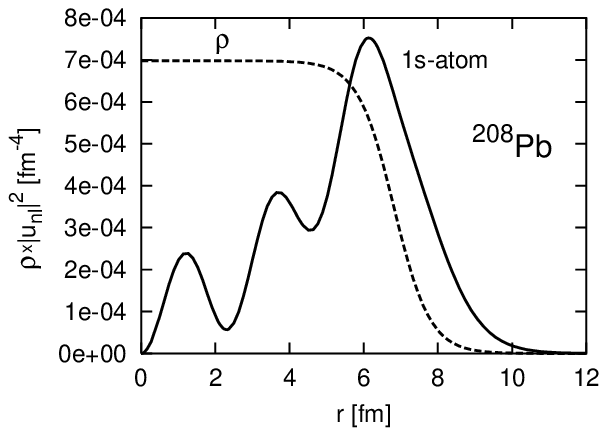}}
\end{center}
%\vspace{-15cm}
\caption {\small As bottom panel of Fig.~\protect\ref{fig:pb7i}, but
now for the deeply bound 1s atomic level, not yet detected.}
\label{fig:pb1s}
\end{figure}

As we mentioned in the introduction, the overlap of the
$K^-$-wave-function and the nucleus is bigger for these low-lying
atomic states than for those accessible via the atomic cascade (see
Fig.~\ref{fig:pb1s}), and thus the precise determination of their
binding energies and widths would provide valuable details of the
antikaon-nucleus interaction at threshold. Thus, several nuclear
reactions have been already suggested to detect them~\cite{zaki00}
and~\cite{FG00}. A rough comparison of the typical values in
Fig.~\ref{fig:pb1s} and those in the bottom panel of
Fig.~\ref{fig:pb7i} helps us to understand why the typical
energy-shifts and widths are of the order of the MeV for deep atomic
states while the magnitude of those  for
high-lying (measured) atomic states was the KeV. The minima of the
1s-atomic modulus squared wave function in Fig.~\ref{fig:pb1s} hint
the existence of deeper bound states, as it was discussed in
Ref.~\cite{FG99}. Those states will be the subject of the next section.

\begin{table}
\begin{center}
\begin{tabular}{l|c|ll|ll|ll|ll}\hline\tstrut
\ 
&\multicolumn{9}{c}{$K^-$ nuclear states} 
\\\hline\tstrut
 \ Nucleus &Nuclear level   &\multicolumn{2}{c|} {(1)}
 &\multicolumn{2}{c|} {(1m)}
 &\multicolumn{2}{c|} {(2)}
 &\multicolumn{2}{c} {(2DD)}
 \\\hline\tstrut 
& $N l$    &~~~$B$ &~$\Gamma$ 
 &~~~$B$ &~~$\Gamma$ 
 &~~~$B$ &~~$\Gamma$ 
 &~~~$B$ &~$\Gamma$ 
 \\
\hline\tstrut
$^{12}$C 
 &0s   &$-$8.64    &92.1&$-$22.1   &43.6&$-$12.2   &183&$-$153     &216  \\
 &0p   &   &  &  &  & &&$-$78.8  &159   \\  
 &0d   &   &  &  &  & &  &$-$6.54  &  108

 \\\hline\tstrut

$^{40}$Ca 
 &0s   &$-$30.6   &105   &$-$44.0   &45.8 &$-$39.6  &213  &$-$92.5   &162  \\
 &0p   &$-$9.71   &91.6 &$-$22.3   &43.3 &$-$14.0  &182  &$-$43.1    &128  

 \\\hline\tstrut 
$^{208}$Pb 
 &0s   &$-$54.1  &107   &$-$66.1  &49.2 &$-$65.8 &196    &$-$197   &205  \\
 &1s   &$-$32.3    &97.6 &$-$43.5  &46.8 &$-$41.4   &177   &$-$160 &181  \\
 &2s   &$-$1.94    &80.7 &$-$14.0 &38.5  &$-$7.19  &149  &$-$110  &152  \\
 &3s   &        &       &       &&& &$-$55.4  &123  \\ 
 &4s   &        &       &       &&& &$-$0.623   &88.0  \\
 &0p   &$-$45.3   &104 &$-$56.9   &48.5  &$-$56.1   &189  &$-$182 &196  \\
 &1p   &$-$18.9    &91.0  &$-$30.1  &44.4   &$-$26.3&165  &$-$138  &168  \\
 &2p   &   &  &  &&& &$-$84.3   & 139  \\
 &3p   &   &  &  &&& &$-$28.7   & 108   \\
 &0d   &$-$35.2    &99.7 &$-$46.4  &47.5  &$-$44.9 &181    &$-$166 &185  \\
 &1d   &$-$4.56   &83.4  &$-$16.1 &40.6    &$-$10.2 &153    &$-$115&155  \\
 &2d   &   &  &  &&& &$-$58.5  &126 \\  
 &0f   &$-$23.9  &94.9  &$-$34.8 &46.1  &$-$32.2 &172  &$-$147  &174  \\
 &1f   &   &  &  &&& &$-$91.0  & 143  \\ 
 &2f   & & &     &&& &$-$33.2 & 113 \\
 &0g   &$-$11.4    &89.4  &$-$22.2&44.3  &$-$18.2  &161  &$-$127   &162  \\
 &1g   &   &  &  &&& &$-$66.8   &  131 \\
 &2g   &   &  &  &&& &$-$8.33  &  98.6  

\\\hline

%\jtstrut
\end{tabular}\\
\end{center}
\caption{ \small Nuclear $K^-$ binding energies ($B$) and widths
($\Gamma$) for different levels ($N,l$), nuclei and optical potentials
($V_{\rm opt}^{\rm (1)}$, $V_{\rm opt}^{\rm (1m)}$, $V_{\rm opt}^{\rm
(2)}$ and $V_{\rm opt}^{\rm (2DD)}$). All units are in MeV. $N$ does
not necessarily denote the number of nodes of the complex wave
function, and just stands for energy ordering.  }
\label{tab:nucl}
\end{table}

  \section{ Nuclear $K^-$ and $\bar K^0$ states}
\label{sec:nucl}
  All four optical potentials defined in the previous sections  also have
  $K^--$nucleus bound levels much deeper and wider than the deep
  atomic states presented in Table~\ref{tab:atom}. We call these
  levels {\it nuclear} states, an enlightening discussion on their
  nature and differences to the atomic states can be found in
  Ref.~\cite{GFB96}. Those states would not exist if the strong
  interaction were switched off. To obtain all the nuclear levels for
  a given optical potential and nucleus, we initially set to zero the
  imaginary part of the optical potential and switch it on gradually
  keeping track at any step of the bound levels. We study three
  different nuclei across the periodic table (carbon, calcium and
  lead).  Results are shown in Table~\ref{tab:nucl}. As can be seen,
  energies and widths depend greatly on the details of the 
  used potential, however, and because of the enormous widths predicted for
  all of them, there exist serious doubts not only on the ability to
  resolve different states but also on their proper existence.

\begin{figure}
\vspace{-1.cm}
\begin{center}                                                                
\leavevmode
\makebox[0cm]{
\epsfysize = 300pt
\epsfbox{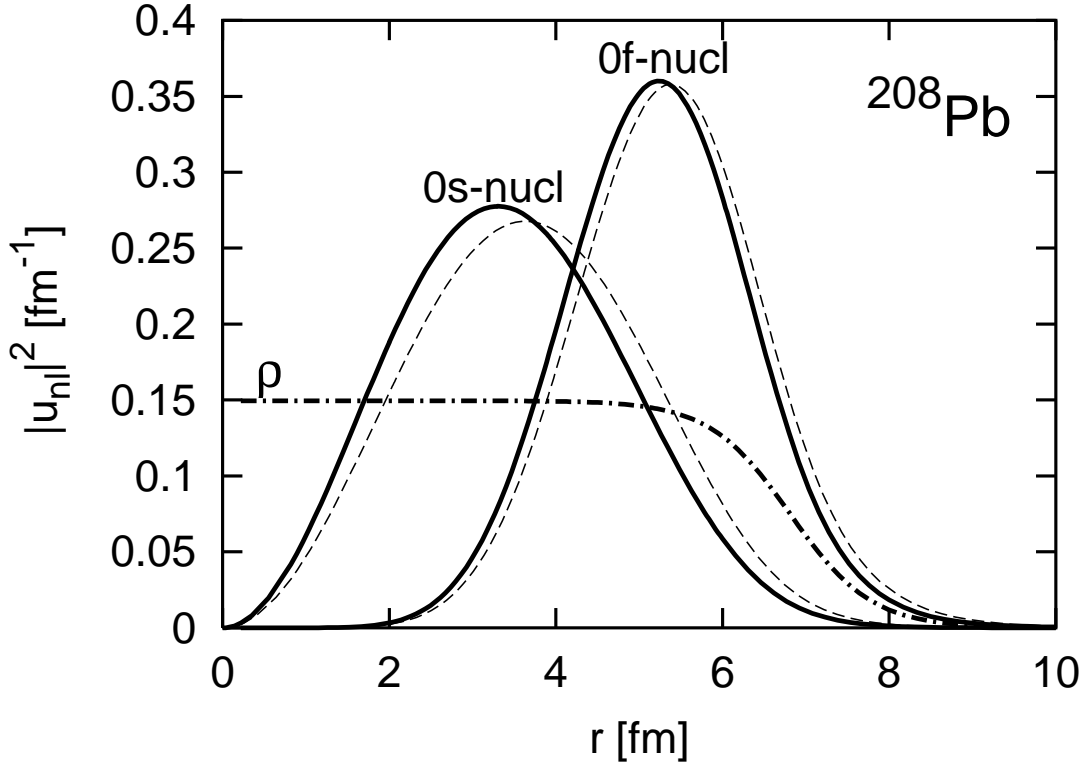}}
\end{center}
%\vspace{-15cm}
\caption {\small Modulus squared of the $K^- -$ (solid) and $\bar
K^0-$ (dashed) $^{208}$Pb reduced radial wave functions for the 0s and
0f nuclear states as functions of $r$. Radial wave functions satisfy
$\int_0^\infty dr~ |u_{nl}(r)|^2 = 1$. We also show the nuclear
density $\rho$ (fm$^{-3}$) in lead. Wave--functions have been obtained
using $V_{\rm opt}^{\rm (1m)}$.}
\label{fig:k0}
\end{figure}

  For the case of the atomic states discussed in the previous
  sections, the net effect of the strong potential was repulsive
  (negative energy shifts), whereas for these nuclear bound states,
  the resulting effect of the strong potential is attractive, as can
  be seen in Table~\ref{tab:k0bar} where we present $\bar K^0-$nuclear
  states. Assuming isospin symmetry and neglecting isovector effects,
  we obtain those neutral antikaon bound states just by switching off
  the electromagnetic potential. Looking at both tables~(\ref{tab:nucl}
  and~\ref{tab:k0bar}), we see that the electromagnetic interaction
  does not affect at all the widths ($\Gamma$) of the deepest
  states. It has some effects on the binding energies ($B$), and, in
  some cases, it is responsible for the existence of some levels for
  $K^-$ and not for the $\bar K^0$ case. In any case, we are certainly
  dealing with a highly non-perturbative (non-linear) scenario.  For
  instance, for all nuclei and levels, the widths are more or less the
  same and only depend significantly on the potential.  This behavior
  is due to the fact that these nuclear wave--functions are totally
  inside of the nucleus (see for instance 0s and 0g nuclear
  $K^--^{208}$Pb levels in Fig.~\ref{fig:pb7i}) where all the optical
  potentials are practically constant and much bigger than 
  the electromagnetic one, as can be seen in
  Fig.~\ref{fig:Vopt1}. Thus, the 
  electromagnetic dynamics does not play a crucial role, as it is
  explicitly shown in Fig.~\ref{fig:k0}, where we compare 
  $K^- -$ and $\bar K^0-$nucleus  wave functions in  $^{208}$Pb.
  As a matter of fact the width of any of those
  levels approximately verifies $\Gamma/2\simeq -~{\rm Im} (V_{\rm
  opt})$ at the center of the nucleus.

  The theoretical potential of Ref.~\cite{RO00} supplemented by the
  phenomenological piece $\delta V^{\rm fit}$ in Eq.~(\ref{eq:V1m}),
  leads both to the best description of the measured kaonic atom data
  and to the narrowest nuclear antikaon states. Indeed, $V_{\rm
  opt}^{\rm (1m)}-$widths are about four or five times smaller than
  the ones predicted by  the empirical $V_{\rm opt}^{\rm (2DD)}$ and
  $V_{\rm opt}^{\rm (2)}$ potentials. Actually, in some cases the $V_{\rm
  opt}^{\rm (1m)}-$interaction predicts states such that $B+\Gamma/2$
  is still negative and thus, they might be interpreted as well defined
  states. 

\begin{table}
\begin{center}
\begin{tabular}{l|c|ll|ll|ll|ll}\hline\tstrut
 & \multicolumn{9}{c}{ $\bar K^0$ nuclear states }
\\\hline\tstrut
 \ Nucleus &Nuclear level   &\multicolumn{2}{c|} {(1)}
 &\multicolumn{2}{c|} {(1m)}
 &\multicolumn{2}{c|} {(2)}
 &\multicolumn{2}{c} {(2DD)}
 \\\hline\tstrut 
& $N l$    &~~~~$B$ &~$\Gamma$ 
 &~~~~$B$ &~~$\Gamma$ 
 &~~~$B$ &~$\Gamma$ 
 &~~~$B$ &~$\Gamma$
 \\
\hline\tstrut

$^{12}$C 
 &0s   &$-$4.76   &91.6 &$-$18.3    &43.4&$-$8.08     &182&$-$149  &216  \\
 &0p   &   &  &   &  &  &  &$-$74.9 &159 \\  
 &0d   &   &  &  &  &  &  &$-$2.83  &  107

 \\\hline\tstrut

$^{40}$Ca 
 &0s   &$-$21.3  &105   &$-$34.7    &45.8&$-$29.9  &212  &$-$83.3   &162  \\
 &0p   &      &  &$-$13.8   &43.0  &$-$4.69  &181 &$-$34.2  &127         

 \\\hline\tstrut 
$^{208}$Pb 
 &0s   &$-$31.1   &107 &$-$43.1   &49.2 &$-$42.4  &195 &$-$173    &205  \\
 &1s   &$-$10.2   &96.4  &$-$21.7   &46.4&$-$18.7   &176  &$-$137  &180  \\
 &2s   &        &       &        &      &      &    &$-$87.9  &152  \\
 &3s   &        &       &        &  &  & &$-$33.5   &122  \\ 
 &0p   &$-$23.4    &103 &$-$35.1 &48.5 &$-$33.7   &188   &$-$160  &195  \\
 &1p   &        &       &$-$8.88    &43.4  &$-$4.07 &164 &$-$115 &167  \\
 &2p   &   &  &                 &  &  &  &$-$62.3  & 138 \\
 &0d   &$-$14.1    &98.6 &$-$25.5  &47.3 &$-$23.1  &179 &$-$143    &184  \\
 &1d   &        &       &       &      &      &   &$-$92.6   &154  \\
 &2d   &   &  &  &  &  &  &$-$36.9  &125 \\  
 &0f   &         &       &$-$14.6 &45.8 &$-$10.9   &170 &$-$125   &173  \\
 &1f   &   &  &  &  &  &  &$-$69.2  & 142 \\ 
 &0g   &        &       &        &      &      &   &$-$106    &162  \\
 &1g   &        &       &      &      &      &     &$-$45.5   &130

\\\hline

%\jtstrut
\end{tabular}\\
\end{center}
\caption{\small Same as in Table~\protect\ref{tab:nucl} for $\bar
K^0-$nuclear states.}
\label{tab:k0bar}
\end{table}

\section{Conclusions}

We have shown that the theoretical potential $V^{(1)}_{\rm opt}$,
recently developed by Ramos and Oset~\cite{RO00} and based on a chiral
model, gives an acceptable description of the observed kaonic atom
states, through the whole periodic table ($\chi^2/dof =
3.8$). Furthermore, it also gives quite reasonable predictions (when
compared to results obtained from other phenomenological potentials
fitted to available data) for the deep kaonic atom states, not yet 
observed. This is noticeable, because it has no free
parameters. Of course, it can be improved by adding to it a small
empirical piece which is fitted to the all available kaonic atom
data. In this way, we have constructed $V_{\rm opt}^{\rm (1m)}$, and
have used it to both quantify the ``deficiencies'' of the
microscopical potential of Ramos and Oset and also to achieve more
reliable predictions for the deeper (atomic and nuclear) bound states
not yet detected. From the first kind of studies, we have concluded
that at low densities, the combined effect of both real and imaginary
parts of the theoretical potential leads to energy shifts more
repulsive than the experimental ones. More quantitative results can be
drawn from Eq.~(\ref{eq:ratio}). Besides, deeply bound atomic state
energies and widths obtained with $V_{\rm opt}^{\rm (1m)}$ confirm the
existence of narrow and separable states, as pointed out in
Ref.~\cite{FG99}, and therefore subject to experimental observation by
means of nuclear reactions. However, there exist appreciable
differences among the predicted widths for these states, when different
potentials are used. Thus, the detection of such states would shed
light on the intricacies of the antikaon behavior inside of a nuclear
medium. 

Finally, we have calculated nuclear antikaon states, for which the
electromagnetic dynamics does not play a crucial role. There, we are in highly
non-perturbative regime and their widths depend dramatically on the
used potential, but they do not depend on the nucleus or
level. However, because of the huge values found for the widths, one
might have some trouble in identifying them as states. In any case,
$V_{\rm opt}^{\rm (1m)}$ leads to the narrowest states, and in some
cases and using $l$-selective nuclear reactions one {\it might}
resolve some states (for instance the ground state in $^{40}$Ca, see
Tables~\ref{tab:nucl} and ~\ref{tab:k0bar}).

To end the paper, we would like to point out, that one should be
cautious when interpreting results and conclusions drawn for the deep
atomic and nuclear states, not yet detected. For instance, though it
is commonly accepted that $p$-wave and isovector corrections to the
optical potential have little effect at the low densities explored by
the available data, it is not necessarily true at the higher densities
explored by deeper bound states.

\section*{Acknowledgments}
We wish to thank E. Friedman, A. Gal, E. Oset, A. Ramos and
L.L. Salcedo for useful communications.  This research was supported
by DGES under contract PB98-1367 and by the Junta de Andaluc\'\i a.

\end{document}